\documentclass[conference]{IEEEtran}
\usepackage{blindtext, graphicx}
\usepackage{booktabs}

\usepackage{cite}

\ifCLASSINFOpdf
\else
\fi
\usepackage{url}

\begin{document}
%
\title{Dynamically Improving Branch\\Prediction Accuracy Between Contexts}

\author{\IEEEauthorblockN{Adam Auten}
\IEEEauthorblockA{Department of Electrical and\\Computer Engineering\\
University of Illinois Urbana-Champaign\\
auten2@illinois.edu}
\and
\IEEEauthorblockN{Tanishq Dubey}
\IEEEauthorblockA{Department of Electrical and\\Computer Engineering\\
University of Illinois Urbana-Champaign\\
tdubey3@illinois.edu}
\and
\IEEEauthorblockN{Rohan Mathur}
\IEEEauthorblockA{Department of Electrical and\\Computer Engineering\\
University of Illinois Urbana-Champaign\\
rmathur2@illinois.edu}}

\maketitle

\begin{abstract}
Branch prediction is a standard feature in most processors, significantly improving the run time of programs by allowing a processor to predict the direction of a branch before it has been evaluated. Current branch prediction methods can achieve excellent prediction accuracy through global tables, various hashing methods, and even machine learning techniques such as SVMs or neural networks. Such designs, however, may lose effectiveness when attempting to predict across context switches in the operating system. Such a scenario may lead to destructive interference between contexts, therefore reducing overall predictor accuracy. To solve this problem, we propose a novel scheme for deciding whether a context switch produces destructive or constructive interference. First, we present evidence that shows that destructive interference can have a significant negative impact on prediction accuracy. Second, we present an extensible framework that keeps track of context switches and prediction accuracy to improve overall accuracy. Experimental results show that this framework effectively reduces the effect of destructive interference on branch prediction.
\end{abstract}


%
\IEEEpeerreviewmaketitle

\section{Introduction}
Processors are using speculative techniques more and more to increase the amount of instruction level parallelism that occurs. This is evident in current cache designs, where data is prefetched from far memory before actual values are ready to be used. Prediction employs the same methodology in order to reduce wasted CPU cycles waiting for a branch to be evaluated. Rather than wait for the branch direction to be evaluated, a prediction is made on the direction of the branch. Correct predictions are rewarded with continuous execution, however, incorrect predictions are given penalized by forcing a processor to employ some sort of backtracking method, thus wasting cycles correcting its error. The deeper the processor pipeline, the more serious this penalty is, hence motivating the need for higher accuracy predictors, usually through radical new designs. We take a different approach -- one that extends current predictor schemes -- by improving the accuracy of predictors between contexts.

Our  work builds on the observation that as a processor switches between the contexts, the state of the branch predictor may destructively interfere with the predictions of the upcoming context. It then becomes natural to ask if we can reduce the effect of these context switches by learning when to clear, or remove, prior history of a branch predictor.

We propose a framework that uses multiple counters and other bookkeeping data to efficiently track and analyze branch predictor behavior across context switches. Our framework not only gives the branch predictor more data to operate on, but can also consider the effects of a context switch through longer histories.

This paper makes the following contributions:
    \begin{itemize}
        \item We provide insights into the behavior of branch predictors with respect to context switches, and show that destructive interference is an issue that can lead to performance degradation.
        \item We introduce the context switch accuracy framework (CSAF), the first framework to use data for contexts as input to a branch predictors, and show that in general cases it can improve upon existing predictor accuracy. For a single core ARMv7 processor running a standard Linux 3.13 kernel with 1ms kernel tick rate simulated in the GEM5 framework, the CSAF improves misprediction on 7 out of 11 tested benchmarks in a predefiend workload.
        \item We explain why the CSAF introduces interesting new ideas for future research.
    \end{itemize}

\section{Related Work}
\subsection{Dynamic Branch Prediction}
Dynamic Branch Prediction has been a well researched topic in recent decades, with modern advances focusing on the improvement, refinement, or various schemes of the two level scheme described by Yeh et al. \cite{twolevel} This method uses a history table full of saturating counters that is indexed using some hash of the branch address. The action of the prediction is based on the current state of the counter, which is updated once the outcome of the branch is evaluated. The problem here is that this scheme, and variations of this scheme, suffer from a few basic problems.

The first of these problems is the aliasing of branch history addresses, and while there have been advances \cite{bimode,combining} to reduce aliasing, the prediction method remains the same, and thus other problems remain unaddressed. 

In addition, history length is a limiting factor in these predictors, first simply by the number of entries that can be stored in the history table and secondly the size of each entry itself. Generous improvements in hardware technology are allowing for larger tables simply through brute force. Table entry sizes are also limited as a function of the number of entries, and as the information in the entry decreases, the less information the predictor has to act upon, including the exclusion of context data from history table entries.

\subsection{Effects of context switching}
Some work has been done on analyzing the effects of context switching and its effect on a running application or overall workload. This research analyzes the performance of an application during a specific workload, however, these papers usually focus on cache performance, with a tangential focus on branch prediction \cite{susceptibility,effect}. They show that the effects of context switching can be diminished in memory through clever cache size manipulations or optimizations based on the expected workload, demonstrating that context switches should not be regarded as trivial to CPU workloads.

Indeed, work has also been done to analyze the effects of context switching on branch prediction. Such research suggests that context switching does not have the expected significant effect on branch prediction accuracy as one would expect, but rather, most accuracies reach a steady state when using realistic time slices for contexts \cite{michele}. However, this is countered by other works which state that in other, nontrivial, workloads, certain branch predictors, such as the two level scheme previously discussed, may be susceptible to accuracy loss when switching contexts \cite{dynamic}.

Other work has also been done to develop branch predictor models that incorporate context data in order to improve prediction accuracy \cite{hybrid}. Results from these works demonstrate that using context data, appreciable results were obtained from these schemes. However, the downside remains that these schemes present wholly new methods for branch prediction, instead of building on top of existing implementations that have already been shown to perform well.

\section{Motivation}
\subsection{Lack of predictors that use contextual data}
When looking at the most common branch prediction schemes, we can see that they are mostly based off the work of Yeh and Patt with their standard two level design. Of course, systems built off this scheme generally exclude the idea of contexts, with most of the data for the predictor coming from current PC address possibly combined with previous accuracy or PC data. This complete oversight of context data means that in thee case where contexts do play a large role in workload performance, the branch predictor will not know how the context data plays a role in the prediction it is about to make. Due to this lack of data, we feel that a gap has been created in the branch prediction realm that completely ignores the fact that in modern computing, contexts switches occur very often and also occur across CPUs in the system. This then begs the question of what would the performance of a context aware branch predictor look like and what sort of insight could a context provide to a prediction scheme.

\subsection{Impact of context switching}
\begin{figure}[h]
\centering
\includegraphics[width=0.5\textwidth]{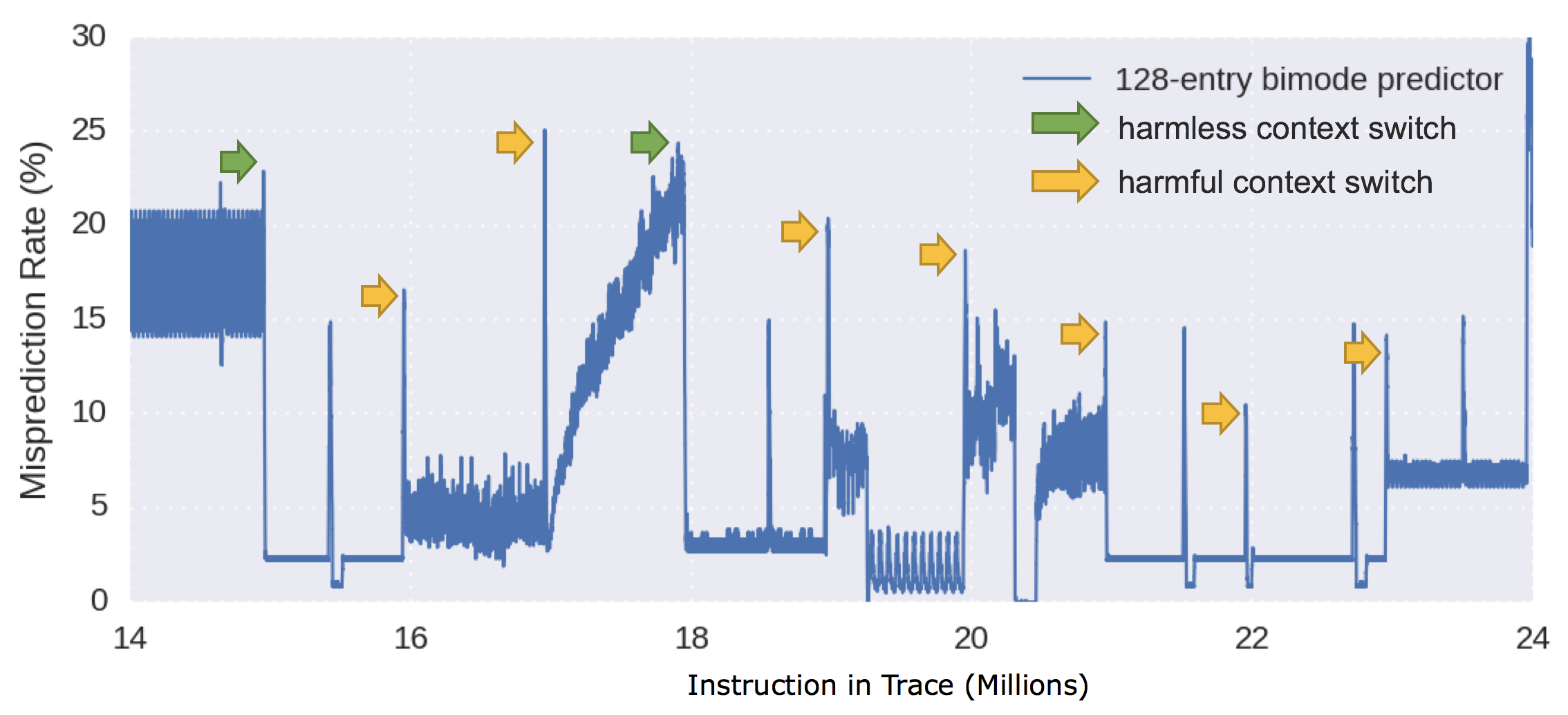}
\caption{Branch misprediction rate vs. time for a multi-process benchmark}
\label{fig:context-trace}
\end{figure}

\begin{figure}[h]
\centering
\includegraphics[width=0.5\textwidth]{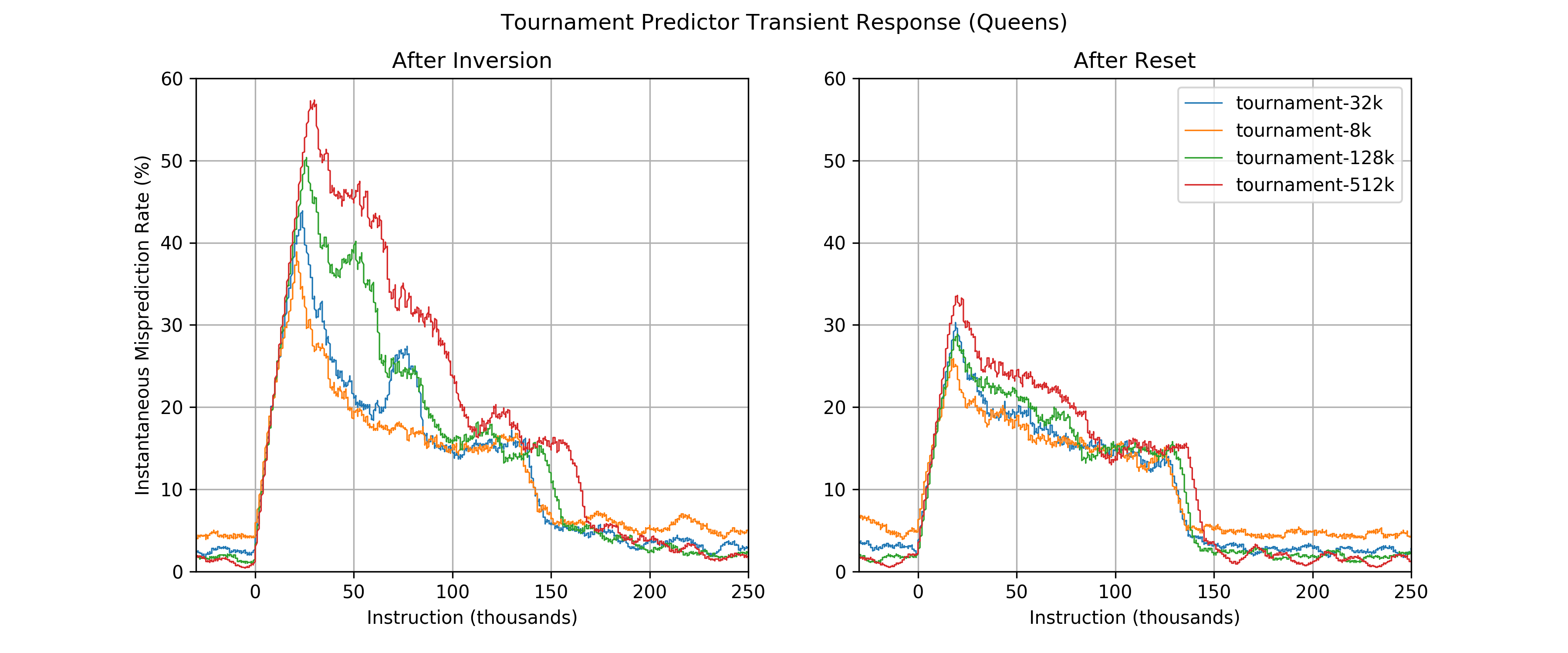}
\caption{Transient response for Tournament branch predictors in response to worst-case aliasing and reset}
\label{fig:tourn}
\end{figure}

In order to further justify the work needed to implement the CSAF, we needed to quantify the amount of impact a context switch has on branch prediction accuracy. For this, a representative baseline workload was constructed to run on the GEM5 simulator with with branch prediction accuracy was recorded. Specifically, a single core ARMv7 processor running a Linux kernel modified to have a context switch every millisecond, ran a workload consisting of eight benchmarks. With this workload, mispredict rate was recorded and graphed. The results can be seen in Figure \ref{fig:context-trace}. As is demonstrated in the graph, there is a significant spike in the mispredict rate at every context switch. These spikes average around 200,000 cycles in length before reaching a steady state, which then leaves room for improvement. In addition to this simulated workload,  a secondary, “worst-case” test was done in order to see what the worst possible context switch might look like. In order to simulate this, two methods were used. First, every ten thousand cycles, all branch prediction history was inverting, meaning every taken was set to not-taken, and every not-taken was set to taken. With this scenario, which is visualized in Figure \ref{fig:tourn}, it can be seen that the mispredict rate spikes to 60\%, with mispredict rate spikes decreasing with predictor size. In the second scenario, the entire history was not inverted, but rather reset to the default predictor state. Similar results were seen here, with mispredict rates spiking to nearly 40\% and decreasing linearly with predictor size. All in all, it is clear that context switching does have an effect on the accuracy of the predictor, leaving space to improve the predictor. 

\section{Description}
When analyzing destructive behavior between contexts when using two-level adaptive branch prediction schemes, we identified that the main source of initial misprediction stems from saturation counters that have been flipped since the last time the context ran in its time slice. The error arises when the new context gets switched in, and begins executing, using the branch history for a previous context's execution patterns, instead of its own. This insight into the poor behavior of branch predictors immediately after a context switch was one of the guiding factors when developing our novel approach to this problem.

Another factor that influenced our design was the fact that a reset of the entire branch predictor's state is often more detrimental than simply using a previous processes branch predictor state, meaning that simply resetting the entire branch predictor upon every context switch would not produce viable results. Only in specific cases of destructive interference is using a previous processes state more detrimental than a reset of branch predictor state, meaning that our design needed to have some adaptive qualities to it.

Using these factors as the primary motivations for our design, we began formulating our design. First, we recognized that resetting the entire predictor state was a very destructive operation, especially in larger predictor tables. During a time slice, we saw that for the benchmarks we ran, there would often only be 15\%-25\% pattern history table (PHT) entry usage by a specific program, with an even fewer number of PHT entries changing their direction (from taken to not taken, or the other way around). Wiping all of the PHT entries regardless of whether they actually changed since the last time that specific program was given a time slice is wasteful. Instead, we chose a more moderate approach, to only wipe the PHT entries that had changed direction (meaning, the PHT entries that changed from taken to not-taken, and vice versa) since the last time the program that is about to run. This ensured that only branch predictor data that remains in the predictor is either reset data, or data that pertains to the upcoming program, both of which are better than the theoretical worst-case state of the predictor.

Secondly, we wanted to control this behavior dynamically, choosing whether to reset the modified PHT entries if the behavior is deemed to be optimal or not. To do this, we took inspiration from PHT tables themselves, by utilizing saturated counters. These counters are each associated with a specific PID to PID transition, and indicate whether the PHT entries that changed since the last run should be reset or not. By only updating these counters when we see behavior that is better or worse by a certain threshold, we remove noise and random blips of branch misprediction data.

In order to update the saturated counters, we need to classify both good and bad behavior. Building off of the insight into what causes mispredictions after a context switch (changed PHT entries), we came up with a metric to measure whether resetting the modified PHT entries was a desirable action or not. By simply tracking the number of PHT changes that take place over the course of a process' time slice, we can compare this number to the previously seen number of PHT entry changes. If it has changed by a certain threshold by getting worse, then we invert the counter.

Updating of the saturated counters and choosing whether to reset the modified PHT entries are both intrinsically linked to a context switch. Because of this, both these actions should take place during each context switch that occurs in the operating system. First, the previous transitions number of PHT changes should be updated with the new value, and the counter should be inverted or not, depending on whether it was a unfavorable or favorable outcome, respectively. After the previous transition's data is updated, we look at the new transitions saturated counter, and reset the PHT entries that were modified since the last time slice if the saturated counters indicate that they should be.

\section{Methodology}
To test this framework out to see how it performs when compared to baseline runs without the framework, we implemented this framework within GEM5, a commonly used system simulation tool.

To hold all of the various counters and keep track of the number of modified PHT entries that changed, we chose to implement this list as a fixed size two dimensional array. This array is indexed by (current\_pid, next\_pid) upon every context switch, with an LRU replacement policy if the array is filled. Every entry in the two dimensional array holds two items. The first piece of data is the number of PHT entry changes that occurred after the last time the corresponding transition was encountered. Secondly, each entry contains a saturated counter, initialized to strongly not taken at the start.

Every context switch, two updates would occur in the two dimensional array. Firstly, the previously seen transition's entry needs to be updated (old\_pid $\rightarrow$ current\_pid). If the previous transitions number of changes in the PHT table was smaller by a certain threshold than the newly found number of changes in the PHT table, then we classify this as worse behavior than what we previously saw. In this case, we should invert the counter, so that the opposite action is taken than previously taken, to see if that method yields better data. If the behavior of the previously taken action is considered to be better than what was stored, then we do nothing. We then update the stored number of modified PHT entries for that particular PID to PID transition entry in the array.

After we update the previously seen transition's entry, we have to decide what to do with the current transition (current\_pid $\rightarrow$ next\_pid). We then look up the new transition’s entry in the table. If the counter indicates taken, then we wipe the modified PHT entries, and continue execution as normal.

If implemented in a real ARM system, one of the biggest hurdles to go from simulation to a real working product is being aware of context switches, along with getting information about process IDs. Luckily, this problem is easily circumvented with a feature present on newer ARM processors, known as ARM Software Thread ID registers \cite{arm}. By monitoring writes these registers, notifying the branch prediction framework of a context switch is very feasible.

\section{Results}
To form a baseline of how context switches effect multiple benchmarks, we simulate context switching between multiple threads. Using the full-system emulation mode \cite{gem5}, we simulated a single core ARM system using the O3\_ARM\_v7a\_3 CPU, running a standard Linux 3.13 kernel with 1 ms kernel tick rate. Both the instantaneous misprediction for the prediction and the average misprediction rate for each process was measured. The instantaneous misprediction rate was measured using a 1000 branch sliding window. Large spikes in the misprediction rate are observed on context switch boundaries.

\begin{figure}[h]
\centering
\includegraphics[width=0.5\textwidth]{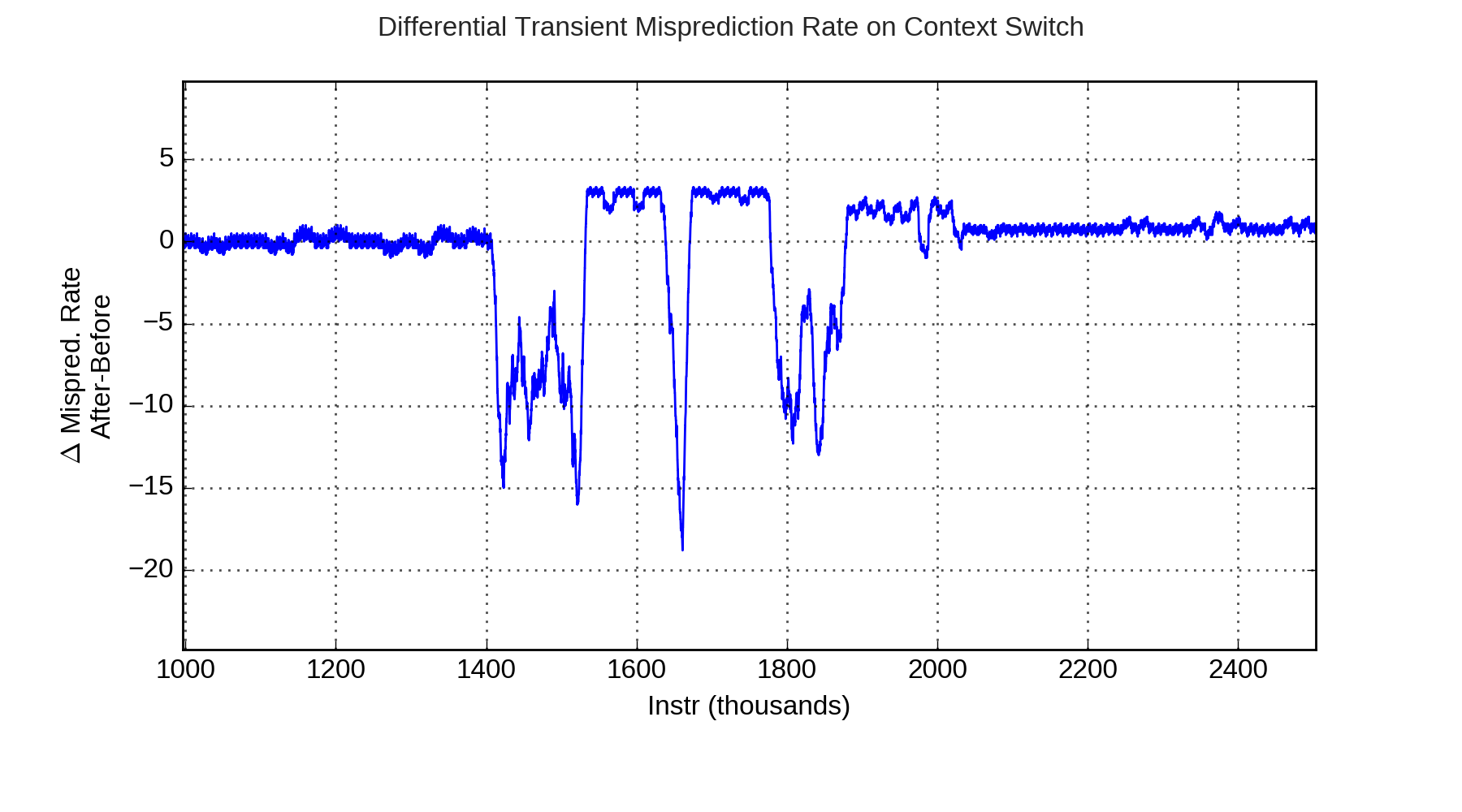}
\caption{Transient response of differential misprediction rate during a context switch}
\label{fig:diff}
\end{figure}

To gauge the effectiveness of our algorithm of conditionally resetting the predictor on destructive context switches, we compared the instantaneous misprediction rate of our algorithm to baseline. Figure \ref{fig:diff} shows the differential misprediction rate in response to a context switch, with the results tabulated in table \ref{tab:results}. The differential rate is calculated by simulating a multi-process benchmark with and without conditional context switch resets. For each time, the instantaneous misprediction rate is subtracted, yielding an instantaneous differential misprediction rate. The large negative spike in mispredicts show that instantaneous mispredictions are reduced at a context switch boundary. 

\begin{figure}[h]
\centering
\includegraphics[width=0.5\textwidth]{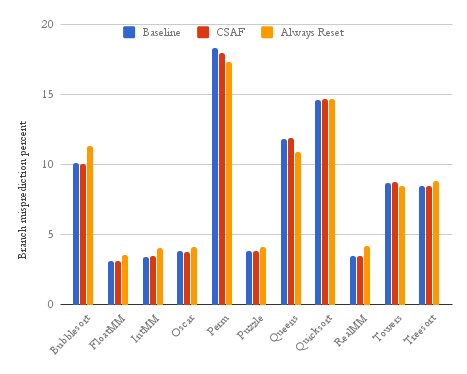}
\includegraphics[width=0.5\textwidth]{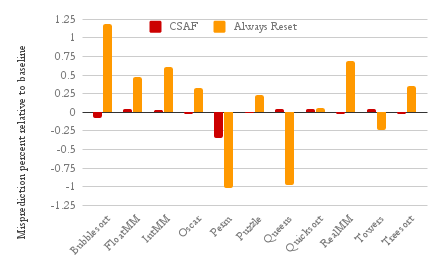}
\caption{Fraction of mispredicted branches per process when all are run together on a single core and context switched with a 1ms time slice}
\label{fig:bar}
\end{figure}

The average misprediction rate of each thread over the course of its execution is shown in figure \ref{fig:bar}. We see that while some benchmarks see a marginal improvement, others are not improved. The small improvement seen is likely due to the infrequency of context switches in our traces (1 every millisecond). Due to limitations in the Linux kernel, we were unable to simulate smaller time slices. 
Interaction between threads determines to a large degree the effectiveness of our algorithm. Misprediction is only improved if there is significant aliasing within the predictor between threads. If the degree of aliasing is small, resetting will do more harm by unnecessarily un-learning any constructive or neutral aliasing. Additionally, if the predictor footprint of a thread is small, resetting the predictor affects the performance of non-contiguously scheduled processes. 

\begin{table}[]
\centering
\caption{Per-Process Misprediction Rates}
\label{tab:results}
\begin{tabular}{@{}lrrr@{}}
\toprule
Benchmark& \multicolumn{3}{c}{Branch Mispredictions (\%)} \\
  \cmidrule{2-4}
  & Baseline & CSAF   & Always Reset \\ 
\midrule

Bubblesort & 10.128   & 10.051 & 11.31        \\
FloatMM    & 3.095    & 3.136  & 3.572        \\
IntMM      & 3.418    & 3.449  & 4.024        \\
Oscar      & 3.811    & 3.786  & 4.136        \\
Perm       & 18.339   & 18.001 & 17.32        \\
Puzzle     & 3.854    & 3.849  & 4.09         \\
Queens     & 11.853   & 11.904 & 10.88        \\
Quicksort  & 14.638   & 14.679 & 14.699       \\
RealMM     & 3.481    & 3.454  & 4.17         \\
Towers     & 8.711    & 8.753  & 8.47         \\
Treesort   & 8.505    & 8.477  & 8.86         \\ 
\bottomrule
\end{tabular}
\end{table}

\section{Future Work}
In our simulations, we used a 128-entry BiMode predictor to enhance the aliasing effects. Future work will involve simulations on larger predictors that are more performant and prevalent in industry. Additionally, instead of inverting the saturating counters on bad behavior, we can use the saturated counters to their full potential by incrementing and decrementing them instead of simply inverting them. In our trials, we found this did not provide much benefit, and that the thresholding was enough to reset when destructive behavior was identified. However, we did not simulate for very long (at a maximum, 500,000,000 CPU cycles), which only encompassed a relatively small number of context switches. Perhaps as a program's behavior got better or worse, using a saturated counter would have allowed for even less unnecessary resets, improving the misprediction rate even further.

When looking at the full branch prediction scheme, most modifications that are made generally change how the table operates or how to fundamentally change the scheme to be more robust. With our findings, we posit a new methodology that could be beneficial to prediction. Instead of modifying the branch predictor by adding hardware components or wholly changing it, we could modify a critical portion of the scheme. Based on recent work by Kraska et al. titled “The Case for Learned Index Structures” \cite{index}, we can see that there is a significant performance in using machine learning to learn the hash function into a map, or the access pattern to an array. In the same lieu, the idea could be applied to branch predictors. Instead of using a static hash function for the predictor table, we can learn the best access pattern to the branch history table. This could potentially significantly improve accuracy as various parameters could be provided to the learning hash function, such as context, to improve performance.


%




\ifCLASSOPTIONcaptionsoff
  \newpage
\fi




\begin{thebibliography}{1}


\bibitem{twolevel}
T. Yeh and Y. Patt, "Two-Level Adaptive Training Branch Prediction", in \emph{Proc. 24th Annual International Symposium in Microarchitecture}, 1991.
\bibitem{bimode}
C. Lee, I. Chen and T. Mudge, "The Bi-Mode Branch Predictor", 1997.
\bibitem{combining}
S. McFarling, "Combining Branch Predictors", in \emph{digital}, Paloƒ Alto, California, 1993.
\bibitem{susceptibility}
W. Hwu and T. Conte, "The susceptibility of programs to context switching", in \emph{IEEE Transactions on Computers}, 1994, pp. 994-1003.
\bibitem{effect}
J. Mogul and A. Borg, "The Effect of Context Switches on Cache Performance", in \emph{digital}, Palo Alto, California, 1990.
\bibitem{michele}
M. Co and K. Skadron, "The Effects of Context Switching on Branch Predictor Performance", 2001.
\bibitem{dynamic}
J. Chen, M. Smith, C. Young and N. Gloy, "An Analysis of Dynamic Branch Prediction Schemes on System Workloads", Philadelphia, 1996.
\bibitem{hybrid}
P. Chang, Y. Patt and M. Evers, "Using Hybrid Branch Predictors to Improve Branch Prediction Accuracy in the Presence of Context Switches", Philadelphia, 1996.
\bibitem{arm}
"ARM Information Center", Infocenter.arm.com, 2017. [Online]. Available: \url{http://infocenter.arm.com/help/index.jsp}. [Accessed: 15- Dec- 2017].
\bibitem{gem5}
"gem5 Full System Simulation — gem5 Tutorial 0.1 documentation", Learning.gem5.org, 2017. [Online]. Available: \url{http://learning.gem5.org/book/part4/intro.html}. [Accessed: 15- Dec- 2017].
\bibitem{index}
T. Kraska, A. Beutel, E. Chi, J. Dean and N. Polyzotis, "The Case for Learned Index Structures", 2017.

\end{thebibliography}
%

%

\begin{IEEEbiography}[{\includegraphics[width=1in,height=1.25in,clip,keepaspectratio]{picture}}]{John Doe}
\blindtext
\end{IEEEbiography}




\end{document}